\documentclass[%
 reprint,
superscriptaddress,
 amsmath,amssymb,
 aps,
prl,
]{revtex4-1}

\usepackage{amsmath}
\usepackage{mathtools}
\usepackage{graphicx}
\usepackage{dcolumn}
\usepackage{bm}
\usepackage{xcolor}
\usepackage{hyperref}
\usepackage{ulem}

\begin{document}

\preprint{APS/123-QED}

\title{A compounded random walk for space-fractional diffusion on finite domains}

\author{Christopher N. Angstmann}
\affiliation{%
	School of Mathematics and Statistics, University of New South Wales, Sydney NSW 2052 Australia
}%

\author{Daniel S. Han}
\email{daniel.han@unsw.edu.au}
\affiliation{%
	School of Mathematics and Statistics, University of New South Wales, Sydney NSW 2052 Australia
}%

\author{Bruce I. Henry}
\affiliation{%
	School of Mathematics and Statistics, University of New South Wales, Sydney NSW 2052 Australia
}%

\author{Boris Z. Huang}
\affiliation{%
	School of Mathematics and Statistics, University of New South Wales, Sydney NSW 2052 Australia
}%

\author{Zhuang Xu}
\affiliation{%
	Centre for Cancer Genetic Epidemiology, University of Cambridge, Strangeways Research Laboratory, Worts Causeway, Cambridge, CB1 8RN, United Kingdom
}%

\date{\today}

\begin{abstract}
	
	We formulate a compounded random walk that is physically well defined on both finite and infinite domains, and samples space-dependent forces throughout jumps. 
	The governing evolution equation for the walk limits to a space-fractional Fokker-Planck equation valid on bounded domains, and recovers the well known superdiffusive space-fractional diffusion equation on infinite domains.
    We describe methods for numerical approximation and Monte Carlo simulations and demonstrate excellent correspondence with analytical solutions.
	This compounded random walk, and its associated fractional Fokker-Planck equation, provides a major advance for modeling space-fractional diffusion through potential fields and on finite domains.
	
\end{abstract}

\maketitle

\section{Introduction}

Many physical processes are known to exhibit anomalous diffusion where the mean squared displacement (MSD) scales nonlinearly with time.
A standard approach to modeling this is through continuous-time random walks (CTRWs) \cite{Montroll1965,Metzler2000}. 
The CTRW describes a random walk in which the time between jumps and the jump lengths are stochastic. 
In the appropriate continuum and diffusion limit, the CTRW master equation can model superdiffusion \cite{Metzler2000} where the MSD grows faster than linearly with time.

Superdiffusion is of interest across a wide range of physical systems. Examples include: random walks with spatial memory \cite{Metzler2000, Fedotov2018memory}, reaction-diffusion equations for evanescent particles \cite{Abad2010}, the motion of endogenous particles within amoeba \cite{Reverey2015}, the trajectories of intracellular lipid bound vesicles \cite{Fedotov2018memory}, diffusion in magnetic resonance imaging data \cite{magin2013characterization}, electrical propagation in heterogeneous media such as biological tissues \cite{bueno2014fractional}, intrusion of contaminated water through sand \cite{benson2000application,kelly2019fractional}, and in aquifers \cite{yin2020super,sun2020review}, animal foraging strategies \cite{humphries2010environmental},
laser driven dissipative plasmas \cite{liu2008superdiffusion}, bacterial swarms \cite{mukherjee2021anomalous} and disordered quantum systems \cite{mildenberger2007boundary}.

Superdiffusion is traditionally generated by CTRWs where the jump length distributions have heavy tails and unbounded variance, namely L\'evy flights \cite{Metzler2000}. 
The heavy tailed jump lengths of L\'evy flights are considered pathological to modeling applications as boundary conditions become problematic \cite{dybiec2017levy}. In physical systems, both boundary conditions and accounting for potential fields during a flight can be critical.  For example, amoebae, intracellular vesicles, contaminated water molecules and foraging animals feel external potential fields generated by their heterogeneous environment during uni-directional flights. It would be unphysical if such random walks could leap over potential wells. The compounded random walk in this paper addresses the pathological nature of L\'evy flights on bounded domains and in potential fields while maintaining the same defining characteristics on the unbounded domain.

For CTRWs with finite mean waiting times and L\'evy flights for jump lengths, the governing equation limits to the space-fractional diffusion equation on the unbounded domain \cite{compte1996stochastic,Metzler2000}
\begin{equation}
	\dfrac{\partial \rho(x,t)}{\partial t} = -D_{\alpha}(-\nabla^2)^\alpha \rho(x,t),
	\label{eq:FractionalDiffusionEquation}
\end{equation}
where  $\alpha \in (0,1]$ is the fractional exponent, $D_{\alpha}$ is the fractional diffusion coefficient and $\rho(x,t)$ is the probability density function (PDF) of finding the particle at position $x$ at time $t$. 
Here the fractional Laplacian $-(-\nabla^2)^\alpha$ is defined via a Riesz derivative, which can be expressed in various ways \cite{bayin2016definition,cai2019riesz,lischke2020fractional}, with the defining property from the Fourier transform
\begin{equation}
	\mathcal{F}\{-(-\nabla^2)^{\alpha}f(x)\}(k) = -|k|^{2\alpha}\mathcal{F}\{f(x)\}(k).
	\label{eq:RieszFourier}
\end{equation}

The particle motion governed by \eqref{eq:FractionalDiffusionEquation} is characterized by a pseudo superdiffusive MSD, $\lim\limits_{\delta\rightarrow2\alpha}\langle |x|^{\delta} (t) \rangle \sim t^\frac{1}{\alpha}$, and $\delta < 2\alpha$ \cite{Metzler2000}. 
While the underlying CTRW stochastic process behind \eqref{eq:FractionalDiffusionEquation} is broadly accepted on infinite domains, there are numerous physical problems on finite domains for which there is no general consensus \cite{lischke2020fractional}. 
Some critical questions arise when considering random walks in finite domains: How can a random walk take its next jump if the jump length exceeds the size of the domain? 
How can such a non-local random walk incorporate possible effects from spatially varying forces, including potential barriers? 
These questions are developing areas of research with pertinent examples being stopped L\'evy flights \cite{Baeumer2018,Kelly2019} and subordinated Brownian motion \cite{Zoia2007,Garbaczewski2019,lischke2020fractional,Sokolov2001,Brockmann2002}. Currently, space-fractional models with the Riesz fractional derivative that incorporate zero flux boundary conditions are unphysical requiring ad-hoc exterior conditions \cite{liu2008superdiffusion,Garbaczewski2019, Brockmann2002}. This is further emphasized by the lack of consistent solutions for the space fractional diffusion equation defined using Riesz fractional derivatives \cite{Monroy2024}.

Alternatively, the fractional order Laplacian can be defined in the spectral sense through the eigenequation \cite{Zoia2007,lischke2020fractional},
\begin{equation}\label{eq:spec}
(-\nabla^2)^\alpha v_n(x) = \lambda_n^{\alpha} v_n(x).
\end{equation}
While the Riesz and spectral fractional Laplacians are equivalent on the infinite domain \cite{Zoia2007,lischke2020fractional}, the spectral fractional Laplacian remains well defined on finite domains. 
In order to use this for modeling purposes it requires a physical framework such as an underlying random walk.

In this paper, we introduce a compounded random walk whose governing equation gives rise to the space-fractional diffusion equation \eqref{eq:FractionalDiffusionEquation} with the fractional Laplacian defined spectrally \eqref{eq:spec} in the diffusion limit. 
The compounding effect will model a process where a particle switches between a waiting process and another random walk on a faster time scale.
The emergence of the spectral Laplacian from the underlying random walk provides a much needed physical interpretation of the space-fractional diffusion equation on bounded domains.
Furthermore, this compounded random walk leads to a space-fractional Fokker-Planck equation on both bounded and unbounded domains with proper path sampling of position dependent forces. 
Our microscopic model is significant for two reasons: Firstly, it provides a method for modeling superdiffusive processes, which are ubiquitous, when a potential is applied to their domain.  Secondly, it provides insight into the underlying dynamics for a system whose behavior is well described by a macroscopic spectral Fokker-Planck equation. We note some related work on this in \cite{Sokolov2001, Brockmann2002}, but the vast bulk of the literature on random walk models for space fractional Fokker-Planck equations has not considered the spectral Fokker-Planck equation.
Our random walk interpretation is naturally suited for Monte Carlo simulation methods, which we verify by comparison to both analytical and numerical solutions. Next, we outline the compounded random walk and its governing equation through a CTRW and find spectral representations for the random walker's probability mass function. 

\section{The compounded CTRW model}
\begin{figure}[h]
	\includegraphics[width=\linewidth]{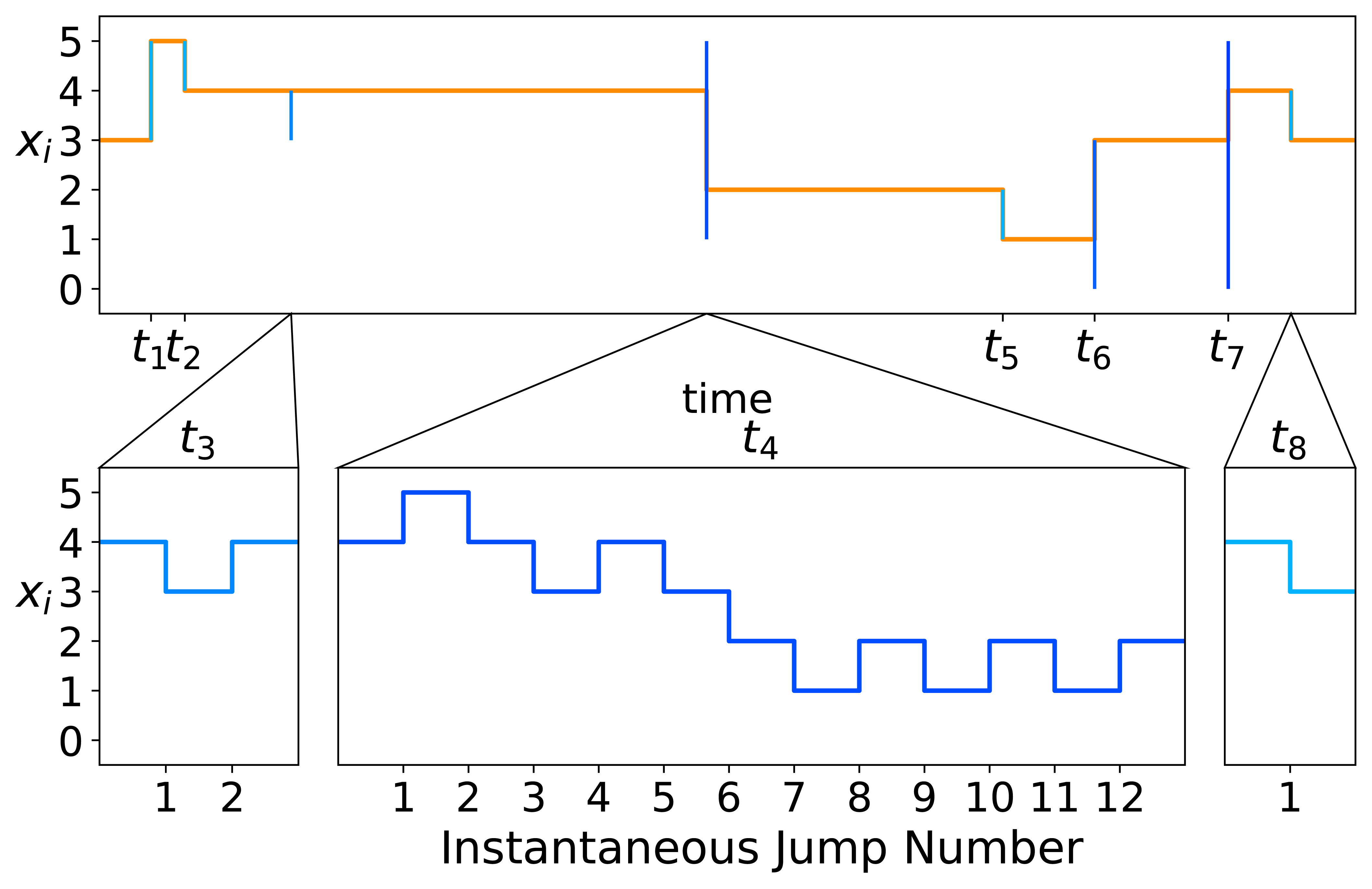}
	\caption{A realization of a single particle performing a compounded CTRW where $\Delta x = 1, a = 0, b = 5$. The top panel shows the lattice site occupancy of the particle where the (orange) horizontal lines show positions and times when the particle is waiting and the (blue) vertical lines show positions sampled by steps during each instantaneous jump. The bottom panels show the $K$ random compounding steps performed by the random walker during jumps at times, $t_3$, $t_4$ and $t_8$.}
	\label{fig:CTRW}
\end{figure}

Consider a particle, performing a CTRW on a lattice, that waits an exponentially distributed amount of time at site $x_i$, with rate $\gamma$, before jumping to a new site $x_j$ governed by a jump distribution $\Lambda(x_j,x_i)$. The lattice is taken to be one-dimensional with spacing $\Delta x = (b-a)/(m-1)$ on a finite interval $[a,b]$ with $m$ lattice points. We denote $\rho(x_i,t)$ as the probability of finding the particle at site $x_i = a+(i-1) \Delta x$ ($1\leq i \leq m$) at time $t$. Due to probability conservation, $\sum_{j} \Lambda (x_j,x_i) = 1$. The master equation for the evolution of $\rho(x_i,t)$ is \cite{Montroll1973,angstmann2015generalized}
\begin{equation}
\dfrac{\partial \rho (x_i,t)}{\partial t} = \gamma \sum_{j=1}^m \Lambda (x_i,x_j) \rho (x_j,t) - \gamma\rho (x_i,t).
\label{eq:master}
\end{equation}
Our aim is to find a jump distribution, $\Lambda(x_i,x_j)$, for a compounded random walk where a jump consists of a random number of Markovian steps. The steps, which make up a single jump, each locally sample the potential which governs its next step, while all occurring instantaneously so that no time passes over the duration of a single jump. The physical interpretation of this is that each jump is constructed from another CTRW with the same potential but with a negligible waiting time in comparison to compounded CTRW. Figure \ref{fig:CTRW} shows an example of a compounded random walk. It is a necessity that after each of these steps the particle remains confined within the domain and thus have finite variance.

We begin by defining a nearest neighbor single step distribution. The particle can step one lattice site to the left or right except if the particle attempts to leave the domain ($1\leq i\leq m$), then it remains at its current site. This can be expressed as the single step transition probability,
\begin{equation}\label{eq:SingleJump}
	\Lambda_1 (x_j,x_i) = 
	r(x_i)\delta_{j,\min(i+1,m)}+l(x_i)\delta_{j,\max(i-1,1)}
\end{equation}
where $\delta_{ji}$ is the Kronecker delta, $r(x_i)$ is the probability to jump to the right at position $x_i$ and $l(x_i)=1-r(x_i)$.
The probability of a step to the right can be related to a potential, $V(x_i)$, and an inverse temperature scale, $\beta$, via a Boltzmann weight \cite{henry2010fractional}
\begin{equation}
r(x_i) = \frac{ e^{-\beta V\left( x_{\min(i+1,m)} \right)}  }{ e^{-\beta V\left( x_{\max(i-1,1)} \right)} + e^{-\beta V\left( x_{\min(i+1,m)} \right)} }.
\label{eq:boltzmann_rightJump}
\end{equation}
For the compounded jump distribution $\Lambda(x_i,x_j)$, in \eqref{eq:master}, we allow the particle to instantaneously step a random number of times $K$ in a single jump such that 
\begin{equation}
	\Lambda(x_j,x_i) = \sum_{k=1}^{\infty} p_K(k) \Lambda_{k}(x_j,x_i),
	\label{eq:CompoundedJumps}
\end{equation}
where $p_K(k) = \text{Prob}\{K=k\}$. The $k$-step distribution is defined recursively using \eqref{eq:SingleJump} as 
\begin{equation}
	\Lambda_k(x_j,x_i) = \sum_{h=1}^{m}\Lambda_1(x_j,x_h)\Lambda_{k-1}(x_h,x_i), \quad k>1.
	\label{eq:Kjumps}
\end{equation}

Substituting the jump distribution \eqref{eq:CompoundedJumps} into the master equation \eqref{eq:master}, we obtain
\begin{equation}
	\begin{split}
	\frac{\partial \rho (x_i,t)}{\partial t} =& \gamma \sum_{j=1}^m \sum_{k=1}^{\infty} p_K(k) \Lambda_{k}(x_i,x_j)  \rho (x_j,t) \\
	&- \gamma\rho (x_i,t).
	\end{split}
	\label{eq:masterCompounded}
\end{equation}
In order to express the right hand side of \eqref{eq:masterCompounded} spectrally, we first write the eigenequation for the single step distribution, 
\begin{equation}
	\sum_{j=1}^m \Lambda_1(x_i,x_j)X_n(x_j) = (1-\mu_n)X_n(x_i),
	\label{eq:SpectralSingleJump}
\end{equation}
which defines $X_n$ as an eigenfunction of the single step distribution with eigenvalue $(1-\mu_n)$. Then the $k$-step distribution will share the eigenfunctions from \eqref{eq:SpectralSingleJump} such that, 
\begin{equation}
	\sum_{j=1}^{m} \Lambda_k (x_i,x_j) X_n (x_j) = (1-\mu_n)^kX_n(x_i).
	\label{eq:SpectralkJump}
\end{equation}
Consequently, the eigenequation for the jump distribution \eqref{eq:CompoundedJumps} is
\begin{equation}
	\sum_{j=1}^m \Lambda(x_i,x_j) X_n(x_j) = G_K(1-\mu_n) X_n(x_i),
	\label{eq:SpectralCompoundedJump}
\end{equation}
where
\begin{equation}
	G_K(z)=\sum_{k=1}^{\infty} p_K(k)z^k,
	\label{eq:ProbabilityGeneratingFunction}
\end{equation} 
is the probability generating function (PGF), given that $p_K(0)= 0$.

Now, we use the spectral property of the jump distribution \eqref{eq:SpectralCompoundedJump} to find the solution of \eqref{eq:masterCompounded} by writing,
\begin{equation}
	\rho(x_i,t) = \sum_{n=1}^{m} c_n X_n(x_i) e^{-\gamma A_{K,n} t},
	\label{eq:GeneralSpectralSolution}
\end{equation}
where for brevity we let $A_{K,n} = 1- G_K(1-\mu_n)$ and $c_n$ is a coefficient determined by the initial condition $\rho(x_i,0)$.
Differentiating \eqref{eq:GeneralSpectralSolution} gives
\begin{equation}
	\frac{\partial \rho (x_i,t)}{\partial t} =-\gamma   \sum_{n=1}^{m} A_{K,n}  c_n X_n(x_i) e^{-\gamma A_{K,n} t}.
	\label{eq:GeneralSpectralRelation}
\end{equation}
Referring back to \eqref{eq:master}, it follows from the equivalence of the right hand sides of \eqref{eq:masterCompounded} and \eqref{eq:GeneralSpectralRelation} that any linear spatial operator of the form
\begin{equation}
	\mathcal{S}\left[f(x_i)\right] = \gamma  \sum_{j=1}^m \left[ \left(  \Lambda(x_i,x_j) -  \delta_{x_i,x_j} \right)  f(x_j) \right],
	\label{eq:NearestNeighborCompoundedLinearOperator}
\end{equation}
has an equivalent spectral representation of the form
\begin{equation}
	\mathcal{S} \left[\rho(x_i,t)\right] = - \gamma   \sum_{n=1}^{m} A_{K,n}  c_n X_n(x_i) e^{-\gamma A_{K,n} t},
	\label{eq:GeneralSpectralLinearOperator}
\end{equation}
where, as above, $A_{K,n}$ can be found from the generating function for the compounded random walk. The validity of \eqref{eq:NearestNeighborCompoundedLinearOperator} and \eqref{eq:GeneralSpectralLinearOperator} relies on the existence of $m$ distinct eigenvalues. This is ensured as the eigenvalue problem for the single step distribution \eqref{eq:SpectralSingleJump} is equivalent to that for a real tridiagonal matrix with strictly positive values. The jump distribution \eqref{eq:SpectralCompoundedJump} having $m$ distinct eigenvalues automatically follows as the PGF \eqref{eq:ProbabilityGeneratingFunction} is monotonically increasing.

Note that this spectral representation is general enough to account for any compounded nearest neighbor random walk in bounded domains and with position dependent potential fields. The master equation for the compounded random walk, \eqref{eq:master} and \eqref{eq:masterCompounded}, can be written in terms of the linear spatial operator \eqref{eq:GeneralSpectralLinearOperator} as
\begin{equation}
	\frac{\partial \rho(x_i,t)}{\partial t} = \mathcal{S}\left[\rho(x_i,t)\right].
	\label{eq:masterCompoundedSpectral}
\end{equation}

To recover the spectral Laplacian \eqref{eq:spec} from the discrete space random walk governed by \eqref{eq:masterCompoundedSpectral}, we choose a specific compounding distribution. As each of the steps have finite variance, to recover the superdiffusive properties of the random walk, we mandate that the distribution of the number of steps itself have infinite mean. For simplicity, we will use the Sibuya distribution defined as \cite{sibuya1979generalized,sibuya1981generalized}
\begin{equation}
	p_K(k) = \binom{\alpha}{k} (-1)^{k+1},
	\label{eq:SibuyaDistribution}
\end{equation}
where $0<\alpha\leq1$, we find the PGF \eqref{eq:ProbabilityGeneratingFunction} gives
\begin{equation}
	G_K(z) = 1- (1-z)^{\alpha}.
	\label{eq:PGFSibuya}
\end{equation}
The compounded random walk with the Sibuya distribution represents a walk where the number of steps within a jump follows a modified Bernoulli process where the probability of stopping on the $k$\textsuperscript{th} step, conditional on not stopping before, is $\alpha/k$. While the Sibuya distribution was used to model power-law waiting times in discrete time random walks \cite{angstmann2015discrete}, here we use it to recover the power-law jump lengths as a sum of local steps, when the domain is unbounded.

Substituting the PGF of the Sibuya distribution \eqref{eq:PGFSibuya} into the spectral representation of the linear spatial operator for the compounded random walk \eqref{eq:GeneralSpectralLinearOperator}, we obtain
\begin{equation}
	\mathcal{S}\left[\rho(x_i,t)\right]  = - \gamma   \sum_{n=1}^{m} \mu_n^{\alpha}  c_n X_n(x_i) e^{-\gamma \mu_n^\alpha t}.
	\label{eq:GeneralSpectralSibuya}
\end{equation}
The linear operator \eqref{eq:GeneralSpectralSibuya} for the unbiased case where $r(x_i) = 1/2$ is a discrete space operator analogous to the spectral fractional Laplacian in \eqref{eq:spec}. 
Next, we take the diffusion limit of \eqref{eq:masterCompoundedSpectral} to obtain the fractional diffusion equation \eqref{eq:FractionalDiffusionEquation}, where the linear spatial operator \eqref{eq:GeneralSpectralSibuya} reverts to the Riesz derivative \eqref{eq:RieszFourier} on unbounded domains. In addition, we show that the governing equation  \eqref{eq:masterCompoundedSpectral} gives rise to a space-fractional Fokker-Planck equation with a spectrally defined operator on both bounded and unbounded domains.

\section{The diffusion limit} 
Extending $X_n(x_i)$ to continuous arguments, $X_n(x)$, so that we can take a Taylor series, we substitute \eqref{eq:SingleJump} and \eqref{eq:boltzmann_rightJump} into \eqref{eq:SpectralSingleJump}.
The Taylor series of $X_n(x\pm\Delta x)$ about $\Delta x = 0$ gives,
\begin{equation}
	\frac{\Delta x^2}{2} \left[ 2 \frac{d}{dx}\left(\beta\frac{dV}{dx}X_n\right) +\frac{d^2X_n}{dx^2} \right] + O\left(\Delta x^4\right) = - \mu_n X_n.
	\label{eq:FP_spectral}
\end{equation}
Using the Fokker-Planck operator,
\begin{equation}
	\mathcal{L}\left[X_n\right] = 2 \frac{d}{dx}\left(\beta\frac{dV}{dx}X_n\right) +\frac{d^2X_n}{dx^2},
	\label{eq:FPoperator}
\end{equation}
we can rewrite \eqref{eq:FP_spectral} as
\begin{equation}
	\frac{\Delta x^2}{2} \mathcal{L}\left[X_n\right] + O\left(\Delta x^4\right) = - \mu_n X_n.
	\label{eq:FP_spectral2}
\end{equation}
To obtain boundary conditions, we can substitute \eqref{eq:SingleJump} and \eqref{eq:boltzmann_rightJump} into \eqref{eq:SpectralSingleJump} at the boundary points $x_1 = a$ and $x_m = b$ to obtain,
\begin{equation}
	\Delta x \left[ \beta \frac{dV}{dx} X_n + \frac{1}{2} \frac{dX_n}{dx} \right] \Bigg\vert_{x=a} + O(\Delta x^2) = -\mu_n X_n(a),
	\label{eq:leftBoundary}
\end{equation}
and
\begin{equation}
	-\Delta x \left[ \beta \frac{dV}{dx} X_n + \frac{1}{2} \frac{dX_n}{dx} \right] \Bigg\vert_{x=b} + O(\Delta x^2) = -\mu_n X_n(b).
	\label{eq:rightBoundary}
\end{equation}
As $\Delta x \rightarrow 0$, we see that the eigenfunctions of the Fokker-Planck operator
\begin{equation}
	\mathcal{L}\left[\chi_n\right] = -\lambda_n \chi_n,
	\label{eq:FP_spectral3}
\end{equation}
must equate to the limit of \eqref{eq:FP_spectral2} since $X_n \rightarrow \chi_n$ in the continuum limit. This equivalence between \eqref{eq:FP_spectral2} and \eqref{eq:FP_spectral3} gives the relationship
\begin{equation}
	\lambda_n \chi_n = \lim_{\Delta x\rightarrow0} \frac{2\mu_n}{\Delta x^2} X_n.
	\label{eq:EigenLimit}
\end{equation}
The eigenfunctions $\chi_n$ are restricted by the continuum limit of \eqref{eq:leftBoundary} and \eqref{eq:rightBoundary} which impose Robin boundary conditions
\begin{equation}
	\beta \frac{dV}{dx} \chi_n + \frac{1}{2}\frac{d\chi_n}{dx} = 0,
	\label{eq:continuousEigenfunctionBoundaryCondition}
\end{equation}
for $x=a$ and $x=b$.

Taking the diffusion limit of \eqref{eq:GeneralSpectralSibuya} and using $\mu_n = \Delta x^2 \lambda_n /2$ from \eqref{eq:EigenLimit} as $\Delta x \rightarrow 0$, we obtain
\begin{equation}
	\mathcal{S}[\rho(x,t)] = - D_{\alpha} \sum_{n=0}^{\infty} \lambda_n^{\alpha}  c_n \chi_n(x) e^{-D_{\alpha}  \lambda_n^{\alpha} t},
	\label{eq:ContinuousSpectralOperator}
\end{equation}
where the mean waiting time is defined by $\tau = 1/\gamma$ and
\begin{equation}
	D_{\alpha} = \lim\limits_{\Delta x, \tau \rightarrow 0}  \frac{\Delta x^{2\alpha}}{2^{\alpha}\tau}.
\end{equation}

The master equation \eqref{eq:masterCompoundedSpectral} defined via the linear spatial operator in continuous space \eqref{eq:ContinuousSpectralOperator} becomes,
\begin{equation}
	\frac{\partial \rho(x,t)}{\partial t} = - D_{\alpha} \sum_{n=0}^{\infty} \lambda_n^{\alpha}  c_n \chi_n(x) e^{-D_{\alpha}  \lambda_n^{\alpha} t}.
	\label{eq:GeneralSpectralSibuyaContinuous2}
\end{equation}

\begin{figure}
	\includegraphics[width=\linewidth]{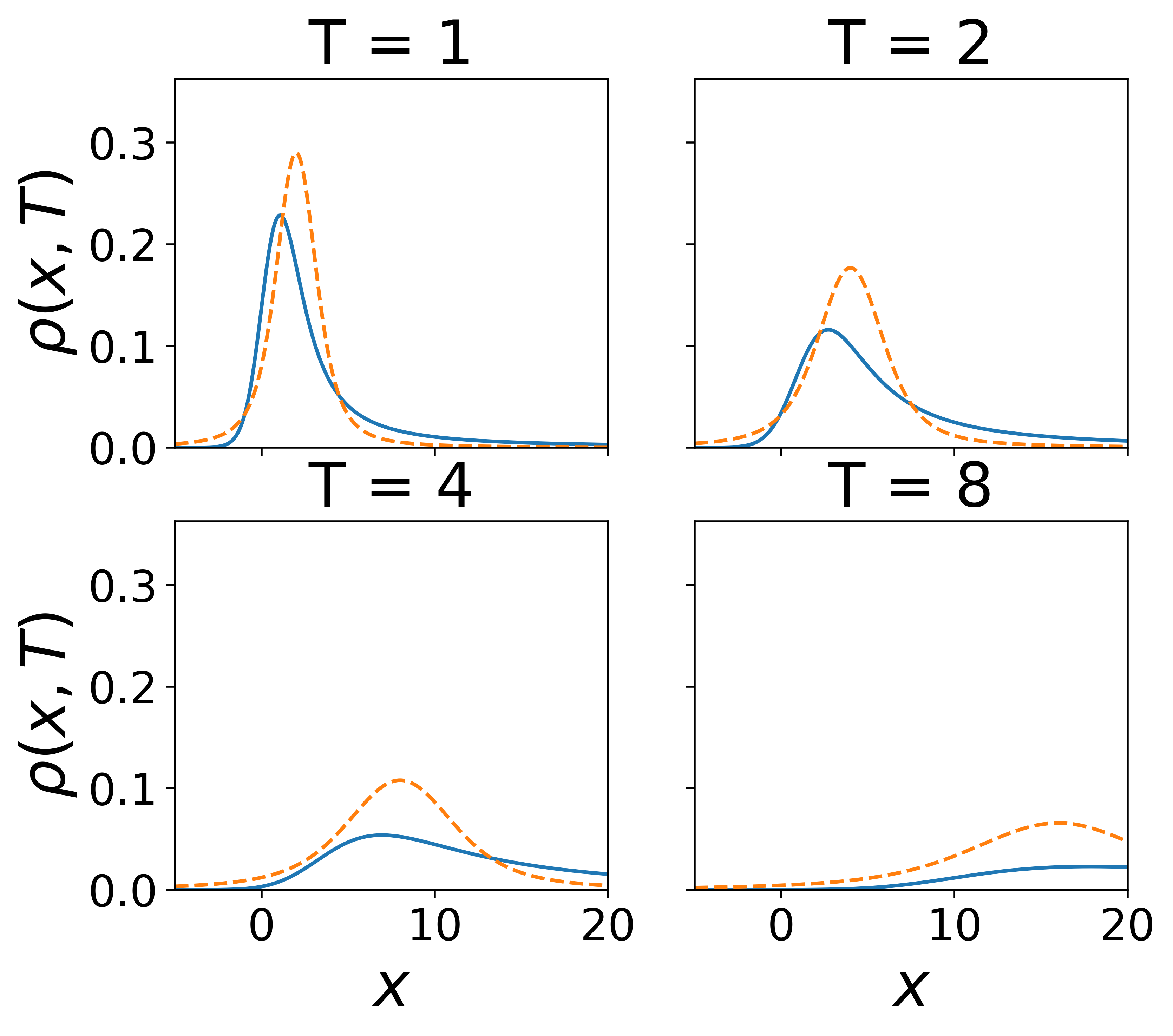}
	\caption{Solutions to the spectral fractional Fokker-Planck equation \eqref{eq:SpectralFractionalFokkerPlanckEquation} (solid blue line) and the fractional advection dispersion equation \eqref{eq:jespersenffpe} (orange dashed line) wtih $D_\alpha = 1$, $v = 1$, $\alpha =0.7$ at time $T = 1$, $2$, $4$, $8$.}
	\label{fig:SFP}
\end{figure}

From \eqref{eq:GeneralSpectralSibuyaContinuous2}, we arrive at a space-fractional Fokker-Planck equation,
\begin{equation}
	\frac{\partial \rho(x,t)}{\partial t} = -D_{\alpha} (-\mathcal{L})^{\alpha}\rho(x,t)
	\label{eq:SpectralFractionalFokkerPlanckEquation}
\end{equation}
where $(-\mathcal{L})^{\alpha}$ is the spectral fractional Fokker-Planck operator. 
From the eigenequation \eqref{eq:FP_spectral3}, the spectral fractional Fokker-Planck operator is defined by
\begin{equation}
	(-\mathcal{L})^{\alpha} \left[ \chi_n \right] = \lambda_n^\alpha \chi_n.
\end{equation}
The significance of \eqref{eq:SpectralFractionalFokkerPlanckEquation} lies in its connection with an underlying compounded random walk whose $K$ random steps in each jump is Sibuya distributed \eqref{eq:SibuyaDistribution}.
Note that the fractional Fokker-Planck operator in \eqref{eq:SpectralFractionalFokkerPlanckEquation} is fundamentally different to previous formulations that arose from L\'evy flights or the Langevin equation on the unbounded domain \cite{fogedby1994levy,fogedby1998levy,jespersen1999levy,Metzler2000}. To demonstrate this, we consider the Fourier transform of \eqref{eq:SpectralFractionalFokkerPlanckEquation}, defined on the unbounded domain
\begin{equation}
	\frac{\partial \hat{\rho}(k,t)}{\partial t} = -D_{\alpha}(|k|^2-ikv)^{\alpha}\hat{\rho}(k,t),
	\label{eq:ourffpe}
\end{equation}
where we have set a constant drift, $2\beta dV/dx = v$ in the Fokker-Planck operator \eqref{eq:FPoperator}. In previous models, the fractional Fokker-Planck equation for constant drift $v$ was defined in the Fourier space by \cite{jespersen1999levy, benson2000application, meerschaert2004finite, shen2008fundamental}
\begin{equation}
	\frac{\partial \hat{\rho}(k,t)}{\partial t} = -(D_\alpha |k|^{2\alpha} - ikv) \hat{\rho}(k,t).
	\label{eq:jespersenffpe}
\end{equation}
Equation \eqref{eq:jespersenffpe} is also known as the fractional advection-dispersion equation \cite{benson2000application,meerschaert2004finite,shen2008fundamental,wang2020fractional}. Comparing \eqref{eq:ourffpe} and \eqref{eq:jespersenffpe}, it is clear that when $v=0$, the two formulations are equivalent and we obtain the space-fractional diffusion equation \eqref{eq:FractionalDiffusionEquation}. This is expected as the Riesz fractional Laplacian is equivalent to the spectral fractional Laplacian for unbounded domains \cite{lischke2020fractional}. 
However, a clear difference between \eqref{eq:ourffpe} and \eqref{eq:jespersenffpe} is that, in \eqref{eq:ourffpe} the advection is felt across the entire path of the underlying random walk and thus encapsulated within the term $(|k|^2+ikv)^{\alpha}$, as opposed to only accounting for it after landing on a site as evident in the separate terms of $D_\alpha |k|^{2\alpha} + ikv$. Our fractional Fokker-Planck equation with the spectral fractional Fokker-Planck operator, \eqref{eq:ourffpe}, is mathematically consistent with \cite{Sokolov2001,Brockmann2002}. Figure \ref{fig:SFP} illustrates the difference between the two governing equations on the unbounded domain at four different times. Unlike the solution to the fractional advection-dispersion equation, \eqref{eq:jespersenffpe}, the solution to the spectral fractional Fokker-Planck equation, \eqref{eq:ourffpe}, does not remain symmetric about the peak. This is due to the compounding of the bias effect in the solution to \eqref{eq:ourffpe}.

For bounded domains, \eqref{eq:SpectralFractionalFokkerPlanckEquation} is identical to the fractional diffusion equation defined via the spectral fractional Laplacian with Neumann boundary conditions when $dV/dx=0$ in \eqref{eq:continuousEigenfunctionBoundaryCondition}. For a spectral fractional Laplacian formulation with Dirichlet boundary conditions, one needs to reformulate \eqref{eq:SingleJump} and take the diffusion limit again, in which case we obtain \eqref{eq:SpectralFractionalFokkerPlanckEquation} with eigenfunctions satisfying boundary conditions $\chi_n(a) = \chi_n(b) = 0$.

Physically, the Dirichlet boundary conditions model the probability distribution for a particle at location $x$ at time $t$ that has never reached the absorbing boundary. This immediately provides a clear use for finding first-passage time distributions for CTRWs making the governing equation well motivated.

\section{Numerical solutions}
The underlying compounded discrete-space random walk for \eqref{eq:SpectralFractionalFokkerPlanckEquation} with zero flux boundary conditions gives a Monte Carlo scheme that can be performed as follows:
\begin{enumerate}
    \item Partition the interval $[a,b]$ into $m-1$ segments of length $\Delta x = (b-a)/(m-1)$ such that there are $m$ nodes approximating position with the first node at $a$. 
    \item Initialize a particle at some initial position $X_0 = x_i$ and set initial time as $t = 0$. 
    \item Draw a random time $\Delta t \sim \text{Exp}(\gamma)$. 
    \item Draw a Sibuya distributed random variable $K$ \cite{Hofert2011}. 
    \item Given $X=x_j$, update $X\leftarrow X+\xi\Delta x$ with the following rules: 
    \begin{enumerate}
    	\item[a)] If $1<j<m$, draw $\xi$ with probabilities $\text{Prob}(\xi=1) = r(X)$ and $\text{Prob}(\xi=-1) = 1-r(X)$;
    	\item[b)] If $j =1$, then  $\text{Prob}(\xi=0) = 1-r(X)$ and $ \text{Prob}(\xi=1) = r(X)$;
    	\item[c)] If $j=m$, then $\text{Prob}(\xi=0) = r(X)$ and $\text{Prob}(\xi=-1) = 1-r(X)$.
    \end{enumerate}
    \item Repeat step (5) an additional $K-1$ times.
    \item Update $t \leftarrow t + \Delta t$. 
    \item Repeat steps (3)-(6) until $t \geq T$, where $T$ is the simulation end time.
\end{enumerate}
The time discretization of \eqref{eq:masterCompounded} gives a numerical scheme for the solution of \eqref{eq:GeneralSpectralSolution}. 

\begin{figure}[h!]
	\includegraphics[width=0.9\linewidth]{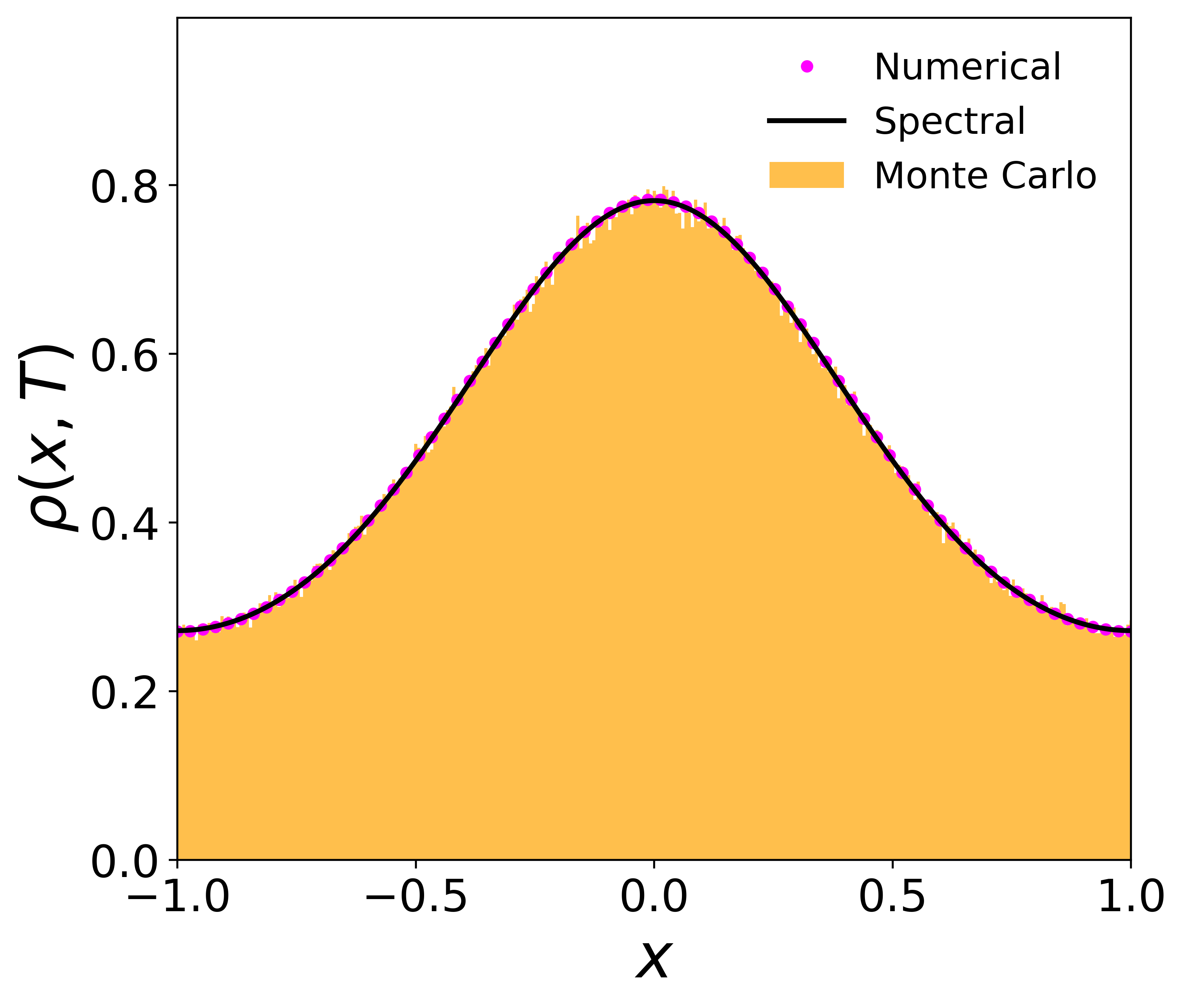}
	\caption{Comparison of solutions between the Monte Carlo simulations (orange), the numerical approximation (purple dots) and the first $1001$ terms of the analytical solution \eqref{eq:spectral_unbiased} (black line) for the space-fractional diffusion equation on bounded domain $[-1,1]$ with reflecting boundaries. The parameters are $\alpha = 0.7$, $N = 10^6$ particles, $\gamma = 10^2$, $m = 301$ and $T = 5$. A subset of data points for the numerical approximation is shown for visual clarity.}
	\label{fig:sim}
\end{figure}

Figure \ref{fig:sim} shows the Monte Carlo simulation, numerical approximation and the exact spectral solution to \eqref{eq:SpectralFractionalFokkerPlanckEquation} with $dV/dx = 0$, on the domain $x \in [-1,1]$ with Neumann boundary conditions and initial condition $\rho(x,0) = \delta(x)$. 
The analytical solution is
\begin{equation}
\rho (x,t) = \dfrac{1}{2} + \sum_{n=1}^{\infty} \exp \left( -D_{\alpha}\left(n \pi\right)^{2\alpha} t\right) \cos\left( n \pi x \right).
\label{eq:spectral_unbiased}
\end{equation}

\begin{figure}[h!]
	\includegraphics[width=\linewidth]{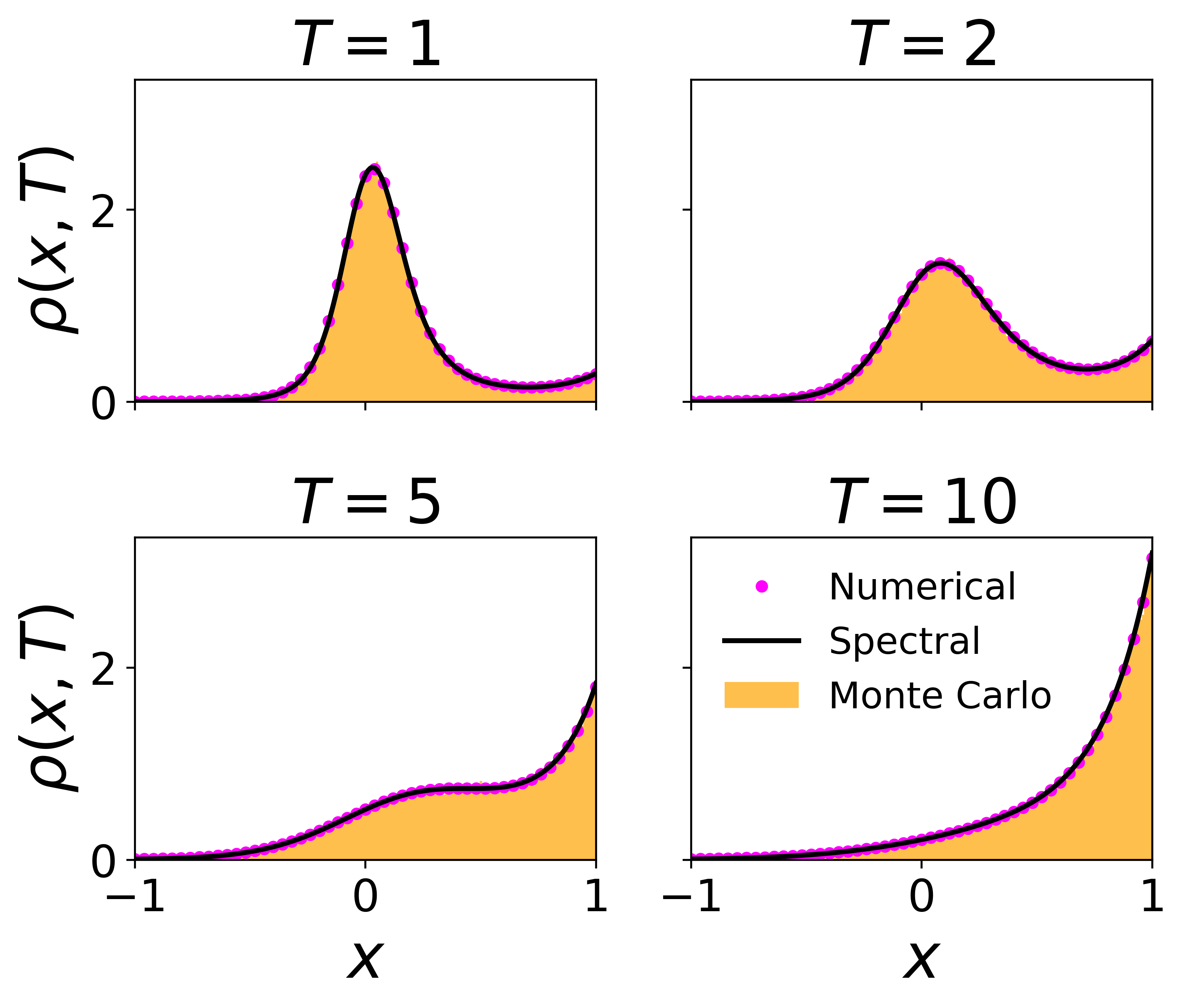}
	\caption{Comparison of solutions between the Monte Carlo simulations (orange), the numerical solution (purple dots) and the first $1001$ terms of the analytical solution \eqref{eq:spectral_biased} (black line) for the space-fractional Fokker-Planck equation on bounded domain $[-1,1]$ with reflecting boundaries, where $\alpha = 0.7$, $N = 10^5$, $\gamma = 50$, $m = 201$ at multiple times, $T = 1$, $2$, $5$ and $10$.}
	\label{fig:bias}
\end{figure}

Now, we extend our results to the biased CTRW with a constant forcing term $\beta V'(x) = -2$. The analytical solution for the governing equation \eqref{eq:SpectralFractionalFokkerPlanckEquation}, given initial condition $\rho(x,0) = \delta (x)$, is 
\begin{equation}
	\begin{split}
		\rho(x,t) = &\dfrac{2e^{4x}}{\sinh(4)} + \sum_{n=1}^\infty  c_n \chi_n(x)  e^ {-D_{\alpha} (4+\omega_n^2)^{\alpha} t} ,
	\end{split}
	\label{eq:spectral_biased}
\end{equation}
where $\omega_n = n\pi/2$
\begin{equation}
c_n =  \dfrac{\omega_n\cos \left( \omega_n \right) + 2 \sin \left(\omega_n \right)}{4+\omega_n^2}
\end{equation}
and
\begin{equation}
	\chi_n(x) = e^{2x}(\omega_n\cos ( \omega_n (x+1) ) + 2 \sin (\omega_n (x+1) )).
\end{equation}
Figure \ref{fig:bias} shows the correspondence between the Monte Carlo simulation, numerical approximation and the truncated spectral solution \eqref{eq:spectral_biased} for the fractional spectral Fokker-Planck operator with constant bias in \eqref{eq:SpectralFractionalFokkerPlanckEquation}. As the solution evolves over time, it is evident that it approaches the Boltzmann distribution, a property not present in other random walk formulations \cite{Sokolov2001}.

\begin{figure}[h]
	\includegraphics[width=\linewidth]{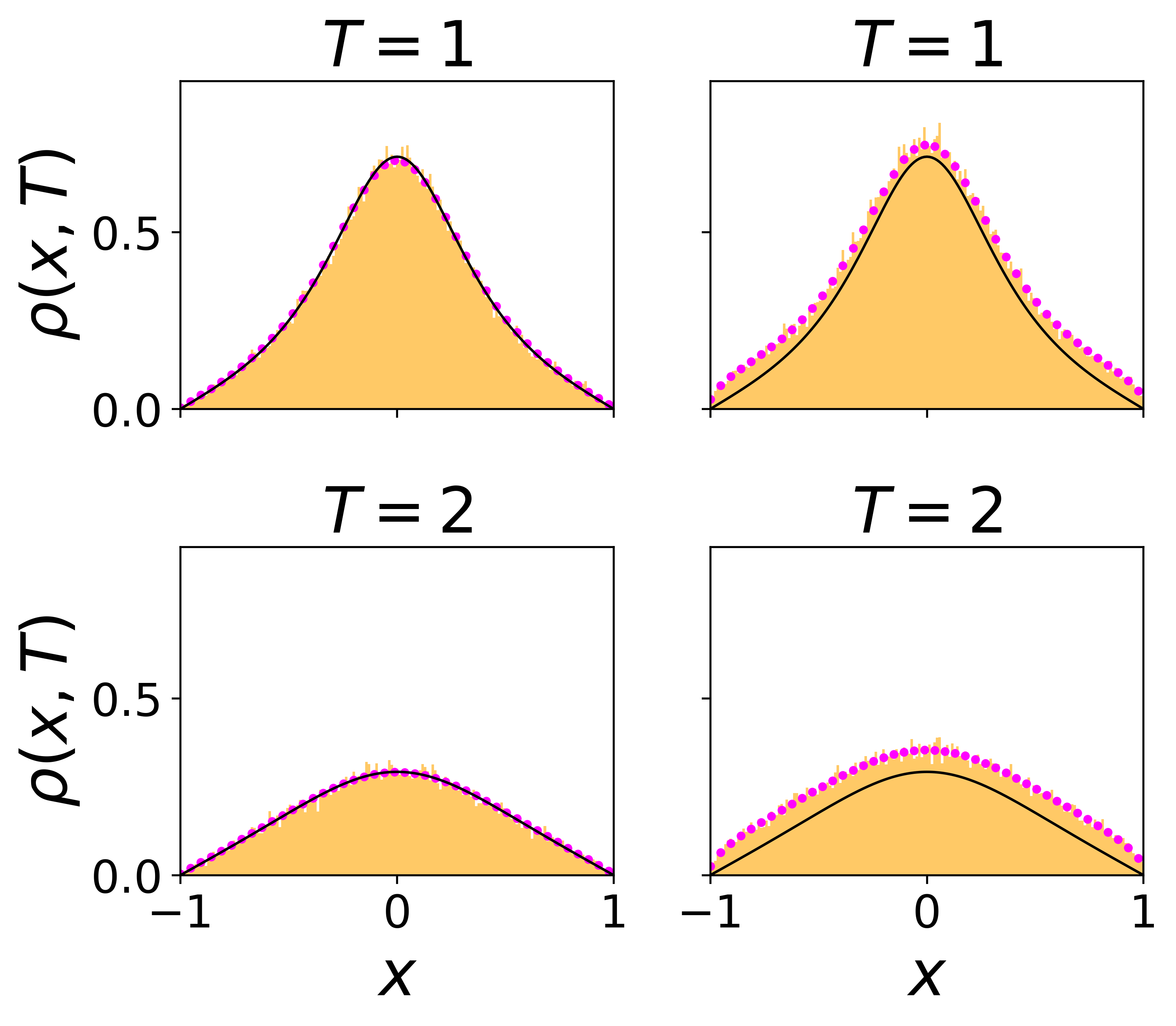}
	\caption{Monte Carlo simulations with $10^5$ particles (orange bars) and their corresponding numerical solutions (magenta dots) of the compounded CTRW in a finite domain $[-1,1]$ at times $T = 1$ (upper panels) and $2$ (lower panels). The case where particles are killed if any point on their path reaches the boundary is on the left and the case where particles are killed if the compounded jump process ends are outside the domain is on the right. The solution to the spectral fractional diffusion equation \eqref{eq:SolDir} is shown as the same solid line on both the left and right panels. Here, $\alpha = 0.5$, the rate $\gamma = 50$ and the number of nodes is $m = 171$.}
	\label{fig:FPD}
\end{figure}

The spectral fractional Fokker-Planck equation may be used to model the survival probability in the first-passage time problem. The solution to \eqref{eq:SpectralFractionalFokkerPlanckEquation} in the case with no bias, absorbing boundaries on the domain $[-1,1]$ and initial condition given by the Dirac delta function $\rho(x,0) = \delta (x)$ is written as
\begin{equation}
    \rho(x,t) = \sum_{n=0}^\infty \exp \left( -D_\alpha \dfrac{(2n+1)^2\pi^2}{4} t \right) \cos \left( \dfrac{(2n+1)\pi x}{2} \right).
    \label{eq:SolDir}
\end{equation}
This solution represents the probability the particle is located at $x$ at time $t$ and has not reached the boundary. Alternatively, if the particles were only absorbed when they had finished a compounded jump outside the domain then the process becomes analogous to killed L\'evy flights on the bounded domain and thus its governing equation gives a finite difference approximation to the space-fractional diffusion equation defined with the Riesz derivative and zero exterior conditions  $\rho(x,t) = 0$ for $x \notin [-1,1]$ \cite{Zoia2007}. Figure \ref{fig:FPD} shows the difference in behavior over time, where it can be seen that there is a significant difference between the case when particles can be absorbed during their compounded jumps (left) and the case where particles can only be absorbed at the end of the compounded jumps (right). This effectively demonstrates that the governing equation with the spectral Laplacian obtained from the compounded CTRW on the bounded domain models the first passage time of a process different to that of CTRW models incorporating L\'evy flights.

\section{Summary and Conclusions}

Our results connect an underlying compounded random walk with the spectral fractional Laplacian in the diffusion limit. This provides a novel and physically valid model to derive the space-fractional diffusion equation on bounded domains. 
Furthermore, this connection naturally generates a spectral fractional Fokker-Planck equation \eqref{eq:SpectralFractionalFokkerPlanckEquation} that is well defined on bounded domains and samples space-dependent forces at every point along a jump. 
The derivation of the fractional Fokker-Planck equation enables a description of random walkers that experience space-dependent forces over the entire path of a single heavy-tailed jump, improving previous formulations \cite{fogedby1994levy,fogedby1998levy,jespersen1999levy}.
In doing so, we open an avenue for models with heavy-tailed jumps where the random walk is sampling the environment throughout the jump and is confined to a finite region in space.
Moreover, through the PGF, \eqref{eq:ProbabilityGeneratingFunction}, new generalized operators based on experimentally observed microscopic mechanisms can be formulated using \eqref{eq:SpectralCompoundedJump}, \eqref{eq:NearestNeighborCompoundedLinearOperator} and \eqref{eq:GeneralSpectralLinearOperator} for models of transport, such as intracellular transport \cite{Reverey2015, Fedotov2018memory}, foraging animals \cite{humphries2010environmental}, and disorderd quantum systems \cite{mildenberger2007boundary}.
The governing equations derived through this approach will have convenient spectral solutions and a natural Monte Carlo simulation scheme.

\begin{acknowledgments}
The authors, C.N.A., B.Z.H. and Z.X., acknowledge financial support by Australian Research Council grant number DP200100345. D.S.H. would like to thank James Norris for useful discussions. The authors would also like to thank an anonymous referee for bringing references \cite{Sokolov2001} and \cite{Brockmann2002} to our attention.
\end{acknowledgments}


\begin{thebibliography}{40}%
	\makeatletter
	\providecommand \@ifxundefined [1]{%
		\@ifx{#1\undefined}
	}%
	\providecommand \@ifnum [1]{%
		\ifnum #1\expandafter \@firstoftwo
		\else \expandafter \@secondoftwo
		\fi
	}%
	\providecommand \@ifx [1]{%
		\ifx #1\expandafter \@firstoftwo
		\else \expandafter \@secondoftwo
		\fi
	}%
	\providecommand \natexlab [1]{#1}%
	\providecommand \enquote  [1]{``#1''}%
	\providecommand \bibnamefont  [1]{#1}%
	\providecommand \bibfnamefont [1]{#1}%
	\providecommand \citenamefont [1]{#1}%
	\providecommand \href@noop [0]{\@secondoftwo}%
	\providecommand \href [0]{\begingroup \@sanitize@url \@href}%
	\providecommand \@href[1]{\@@startlink{#1}\@@href}%
	\providecommand \@@href[1]{\endgroup#1\@@endlink}%
	\providecommand \@sanitize@url [0]{\catcode `\\12\catcode `\$12\catcode
		`\&12\catcode `\#12\catcode `\^12\catcode `\_12\catcode `\%12\relax}%
	\providecommand \@@startlink[1]{}%
	\providecommand \@@endlink[0]{}%
	\providecommand \url  [0]{\begingroup\@sanitize@url \@url }%
	\providecommand \@url [1]{\endgroup\@href {#1}{\urlprefix }}%
	\providecommand \urlprefix  [0]{URL }%
	\providecommand \Eprint [0]{\href }%
	\providecommand \doibase [0]{http://dx.doi.org/}%
	\providecommand \selectlanguage [0]{\@gobble}%
	\providecommand \bibinfo  [0]{\@secondoftwo}%
	\providecommand \bibfield  [0]{\@secondoftwo}%
	\providecommand \translation [1]{[#1]}%
	\providecommand \BibitemOpen [0]{}%
	\providecommand \bibitemStop [0]{}%
	\providecommand \bibitemNoStop [0]{.\EOS\space}%
	\providecommand \EOS [0]{\spacefactor3000\relax}%
	\providecommand \BibitemShut  [1]{\csname bibitem#1\endcsname}%
	\let\auto@bib@innerbib\@empty
	\bibitem [{\citenamefont {Montroll}\ and\ \citenamefont
		{Weiss}(1965)}]{Montroll1965}%
	\BibitemOpen
	\bibfield  {author} {\bibinfo {author} {\bibfnamefont {E.~W.}\ \bibnamefont
			{Montroll}}\ and\ \bibinfo {author} {\bibfnamefont {G.~H.}\ \bibnamefont
			{Weiss}},\ }\href@noop {} {\bibfield  {journal} {\bibinfo  {journal} {Journal
				of Mathematical Physics}\ }\textbf {\bibinfo {volume} {6}},\ \bibinfo {pages}
		{167} (\bibinfo {year} {1965})}\BibitemShut {NoStop}%
	\bibitem [{\citenamefont {Metzler}\ and\ \citenamefont
		{Klafter}(2000)}]{Metzler2000}%
	\BibitemOpen
	\bibfield  {author} {\bibinfo {author} {\bibfnamefont {R.}~\bibnamefont
			{Metzler}}\ and\ \bibinfo {author} {\bibfnamefont {J.}~\bibnamefont
			{Klafter}},\ }\href {\doibase https://doi.org/10.1016/S0370-1573(00)00070-3}
	{\bibfield  {journal} {\bibinfo  {journal} {Physics Reports}\ }\textbf
		{\bibinfo {volume} {339}},\ \bibinfo {pages} {1} (\bibinfo {year}
		{2000})}\BibitemShut {NoStop}%
	\bibitem [{\citenamefont {Fedotov}\ \emph {et~al.}(2018)\citenamefont
		{Fedotov}, \citenamefont {Korabel}, \citenamefont {Waigh}, \citenamefont
		{Han},\ and\ \citenamefont {Allan}}]{Fedotov2018memory}%
	\BibitemOpen
	\bibfield  {author} {\bibinfo {author} {\bibfnamefont {S.}~\bibnamefont
			{Fedotov}}, \bibinfo {author} {\bibfnamefont {N.}~\bibnamefont {Korabel}},
		\bibinfo {author} {\bibfnamefont {T.~A.}\ \bibnamefont {Waigh}}, \bibinfo
		{author} {\bibfnamefont {D.}~\bibnamefont {Han}}, \ and\ \bibinfo {author}
		{\bibfnamefont {V.~J.}\ \bibnamefont {Allan}},\ }\href {\doibase
		10.1103/PhysRevE.98.042136} {\bibfield  {journal} {\bibinfo  {journal} {Phys.
				Rev. E}\ }\textbf {\bibinfo {volume} {98}},\ \bibinfo {pages} {042136}
		(\bibinfo {year} {2018})}\BibitemShut {NoStop}%
	\bibitem [{\citenamefont {Abad}\ \emph {et~al.}(2010)\citenamefont {Abad},
		\citenamefont {BravoYuste},\ and\ \citenamefont {Lindenberg}}]{Abad2010}%
	\BibitemOpen
	\bibfield  {author} {\bibinfo {author} {\bibfnamefont {E.}~\bibnamefont
			{Abad}}, \bibinfo {author} {\bibfnamefont {S.}~\bibnamefont {BravoYuste}}, \
		and\ \bibinfo {author} {\bibfnamefont {K.}~\bibnamefont {Lindenberg}},\
	}\href {\doibase 10.1103/PhysRevE.81.031115} {\bibfield  {journal} {\bibinfo
			{journal} {Phys. Rev. E}\ }\textbf {\bibinfo {volume} {81}},\ \bibinfo
		{pages} {031115} (\bibinfo {year} {2010})}\BibitemShut {NoStop}%
	\bibitem [{\citenamefont {Reverey}\ \emph {et~al.}(2015)\citenamefont
		{Reverey}, \citenamefont {Jeon}, \citenamefont {Bao}, \citenamefont {Leippe},
		\citenamefont {Metzler},\ and\ \citenamefont {Selhuber-Unkel}}]{Reverey2015}%
	\BibitemOpen
	\bibfield  {author} {\bibinfo {author} {\bibfnamefont {J.}~\bibnamefont
			{Reverey}}, \bibinfo {author} {\bibfnamefont {J.-H.}\ \bibnamefont {Jeon}},
		\bibinfo {author} {\bibfnamefont {H.}~\bibnamefont {Bao}}, \bibinfo {author}
		{\bibfnamefont {M.}~\bibnamefont {Leippe}}, \bibinfo {author} {\bibfnamefont
			{R.}~\bibnamefont {Metzler}}, \ and\ \bibinfo {author} {\bibfnamefont
			{C.}~\bibnamefont {Selhuber-Unkel}},\ }\href {\doibase 10.1038/srep11690}
	{\bibfield  {journal} {\bibinfo  {journal} {Scientific Reports}\ }\textbf
		{\bibinfo {volume} {5}},\ \bibinfo {pages} {11690} (\bibinfo {year}
		{2015})}\BibitemShut {NoStop}%
	\bibitem [{\citenamefont {Magin}\ \emph {et~al.}(2013)\citenamefont {Magin},
		\citenamefont {Ingo}, \citenamefont {Colon-Perez}, \citenamefont {Triplett},\
		and\ \citenamefont {Mareci}}]{magin2013characterization}%
	\BibitemOpen
	\bibfield  {author} {\bibinfo {author} {\bibfnamefont {R.~L.}\ \bibnamefont
			{Magin}}, \bibinfo {author} {\bibfnamefont {C.}~\bibnamefont {Ingo}},
		\bibinfo {author} {\bibfnamefont {L.}~\bibnamefont {Colon-Perez}}, \bibinfo
		{author} {\bibfnamefont {W.}~\bibnamefont {Triplett}}, \ and\ \bibinfo
		{author} {\bibfnamefont {T.~H.}\ \bibnamefont {Mareci}},\ }\href@noop {}
	{\bibfield  {journal} {\bibinfo  {journal} {Microporous and Mesoporous
				Materials}\ }\textbf {\bibinfo {volume} {178}},\ \bibinfo {pages} {39}
		(\bibinfo {year} {2013})}\BibitemShut {NoStop}%
	\bibitem [{\citenamefont {Bueno-Orovio}\ \emph {et~al.}(2014)\citenamefont
		{Bueno-Orovio}, \citenamefont {Kay}, \citenamefont {Grau}, \citenamefont
		{Rodriguez},\ and\ \citenamefont {Burrage}}]{bueno2014fractional}%
	\BibitemOpen
	\bibfield  {author} {\bibinfo {author} {\bibfnamefont {A.}~\bibnamefont
			{Bueno-Orovio}}, \bibinfo {author} {\bibfnamefont {D.}~\bibnamefont {Kay}},
		\bibinfo {author} {\bibfnamefont {V.}~\bibnamefont {Grau}}, \bibinfo {author}
		{\bibfnamefont {B.}~\bibnamefont {Rodriguez}}, \ and\ \bibinfo {author}
		{\bibfnamefont {K.}~\bibnamefont {Burrage}},\ }\href@noop {} {\bibfield
		{journal} {\bibinfo  {journal} {Journal of The Royal Society Interface}\
		}\textbf {\bibinfo {volume} {11}},\ \bibinfo {pages} {20140352} (\bibinfo
		{year} {2014})}\BibitemShut {NoStop}%
	\bibitem [{\citenamefont {Benson}\ \emph {et~al.}(2000)\citenamefont {Benson},
		\citenamefont {Wheatcraft},\ and\ \citenamefont
		{Meerschaert}}]{benson2000application}%
	\BibitemOpen
	\bibfield  {author} {\bibinfo {author} {\bibfnamefont {D.~A.}\ \bibnamefont
			{Benson}}, \bibinfo {author} {\bibfnamefont {S.~W.}\ \bibnamefont
			{Wheatcraft}}, \ and\ \bibinfo {author} {\bibfnamefont {M.~M.}\ \bibnamefont
			{Meerschaert}},\ }\href@noop {} {\bibfield  {journal} {\bibinfo  {journal}
			{Water resources research}\ }\textbf {\bibinfo {volume} {36}},\ \bibinfo
		{pages} {1403} (\bibinfo {year} {2000})}\BibitemShut {NoStop}%
	\bibitem [{\citenamefont {Kelly}\ and\ \citenamefont
		{Meerschaert}(2019)}]{kelly2019fractional}%
	\BibitemOpen
	\bibfield  {author} {\bibinfo {author} {\bibfnamefont {J.~F.}\ \bibnamefont
			{Kelly}}\ and\ \bibinfo {author} {\bibfnamefont {M.~M.}\ \bibnamefont
			{Meerschaert}},\ }\href@noop {} {\bibfield  {journal} {\bibinfo  {journal}
			{Application in physics, part B}\ ,\ \bibinfo {pages} {129}} (\bibinfo {year}
		{2019})}\BibitemShut {NoStop}%
	\bibitem [{\citenamefont {Yin}\ \emph {et~al.}(2020)\citenamefont {Yin},
		\citenamefont {Zhang}, \citenamefont {Ma}, \citenamefont {Tick},
		\citenamefont {Bianchi}, \citenamefont {Zheng}, \citenamefont {Wei},
		\citenamefont {Wei},\ and\ \citenamefont {Liu}}]{yin2020super}%
	\BibitemOpen
	\bibfield  {author} {\bibinfo {author} {\bibfnamefont {M.}~\bibnamefont
			{Yin}}, \bibinfo {author} {\bibfnamefont {Y.}~\bibnamefont {Zhang}}, \bibinfo
		{author} {\bibfnamefont {R.}~\bibnamefont {Ma}}, \bibinfo {author}
		{\bibfnamefont {G.~R.}\ \bibnamefont {Tick}}, \bibinfo {author}
		{\bibfnamefont {M.}~\bibnamefont {Bianchi}}, \bibinfo {author} {\bibfnamefont
			{C.}~\bibnamefont {Zheng}}, \bibinfo {author} {\bibfnamefont
			{W.}~\bibnamefont {Wei}}, \bibinfo {author} {\bibfnamefont {S.}~\bibnamefont
			{Wei}}, \ and\ \bibinfo {author} {\bibfnamefont {X.}~\bibnamefont {Liu}},\
	}\href@noop {} {\bibfield  {journal} {\bibinfo  {journal} {Journal of
				Hydrology}\ }\textbf {\bibinfo {volume} {582}},\ \bibinfo {pages} {124515}
		(\bibinfo {year} {2020})}\BibitemShut {NoStop}%
	\bibitem [{\citenamefont {Sun}\ \emph {et~al.}(2020)\citenamefont {Sun},
		\citenamefont {Qiu}, \citenamefont {Wu}, \citenamefont {Niu},\ and\
		\citenamefont {Hu}}]{sun2020review}%
	\BibitemOpen
	\bibfield  {author} {\bibinfo {author} {\bibfnamefont {L.}~\bibnamefont
			{Sun}}, \bibinfo {author} {\bibfnamefont {H.}~\bibnamefont {Qiu}}, \bibinfo
		{author} {\bibfnamefont {C.}~\bibnamefont {Wu}}, \bibinfo {author}
		{\bibfnamefont {J.}~\bibnamefont {Niu}}, \ and\ \bibinfo {author}
		{\bibfnamefont {B.~X.}\ \bibnamefont {Hu}},\ }\href@noop {} {\bibfield
		{journal} {\bibinfo  {journal} {Wiley Interdisciplinary Reviews: Water}\
		}\textbf {\bibinfo {volume} {7}},\ \bibinfo {pages} {e1448} (\bibinfo {year}
		{2020})}\BibitemShut {NoStop}%
	\bibitem [{\citenamefont {Humphries}\ \emph {et~al.}(2010)\citenamefont
		{Humphries}, \citenamefont {Queiroz}, \citenamefont {Dyer}, \citenamefont
		{Pade}, \citenamefont {Musyl}, \citenamefont {Schaefer}, \citenamefont
		{Fuller}, \citenamefont {Brunnschweiler}, \citenamefont {Doyle},
		\citenamefont {Houghton} \emph {et~al.}}]{humphries2010environmental}%
	\BibitemOpen
	\bibfield  {author} {\bibinfo {author} {\bibfnamefont {N.~E.}\ \bibnamefont
			{Humphries}}, \bibinfo {author} {\bibfnamefont {N.}~\bibnamefont {Queiroz}},
		\bibinfo {author} {\bibfnamefont {J.~R.}\ \bibnamefont {Dyer}}, \bibinfo
		{author} {\bibfnamefont {N.~G.}\ \bibnamefont {Pade}}, \bibinfo {author}
		{\bibfnamefont {M.~K.}\ \bibnamefont {Musyl}}, \bibinfo {author}
		{\bibfnamefont {K.~M.}\ \bibnamefont {Schaefer}}, \bibinfo {author}
		{\bibfnamefont {D.~W.}\ \bibnamefont {Fuller}}, \bibinfo {author}
		{\bibfnamefont {J.~M.}\ \bibnamefont {Brunnschweiler}}, \bibinfo {author}
		{\bibfnamefont {T.~K.}\ \bibnamefont {Doyle}}, \bibinfo {author}
		{\bibfnamefont {J.~D.}\ \bibnamefont {Houghton}},  \emph {et~al.},\
	}\href@noop {} {\bibfield  {journal} {\bibinfo  {journal} {Nature}\ }\textbf
		{\bibinfo {volume} {465}},\ \bibinfo {pages} {1066} (\bibinfo {year}
		{2010})}\BibitemShut {NoStop}%
	\bibitem [{\citenamefont {Liu}\ and\ \citenamefont
		{Goree}(2008)}]{liu2008superdiffusion}%
	\BibitemOpen
	\bibfield  {author} {\bibinfo {author} {\bibfnamefont {B.}~\bibnamefont
			{Liu}}\ and\ \bibinfo {author} {\bibfnamefont {J.}~\bibnamefont {Goree}},\
	}\href@noop {} {\bibfield  {journal} {\bibinfo  {journal} {Phys. Rev. Lett.}\
		}\textbf {\bibinfo {volume} {100}},\ \bibinfo {pages} {055003} (\bibinfo
		{year} {2008})}\BibitemShut {NoStop}%
	\bibitem [{\citenamefont {Mukherjee}\ \emph {et~al.}(2021)\citenamefont
		{Mukherjee}, \citenamefont {Singh}, \citenamefont {James},\ and\
		\citenamefont {Ray}}]{mukherjee2021anomalous}%
	\BibitemOpen
	\bibfield  {author} {\bibinfo {author} {\bibfnamefont {S.}~\bibnamefont
			{Mukherjee}}, \bibinfo {author} {\bibfnamefont {R.~K.}\ \bibnamefont
			{Singh}}, \bibinfo {author} {\bibfnamefont {M.}~\bibnamefont {James}}, \ and\
		\bibinfo {author} {\bibfnamefont {S.~S.}\ \bibnamefont {Ray}},\ }\href@noop
	{} {\bibfield  {journal} {\bibinfo  {journal} {Phys. Rev. Lett.}\ }\textbf
		{\bibinfo {volume} {127}},\ \bibinfo {pages} {118001} (\bibinfo {year}
		{2021})}\BibitemShut {NoStop}%
	\bibitem [{\citenamefont {Mildenberger}\ \emph {et~al.}(2007)\citenamefont
		{Mildenberger}, \citenamefont {Subramaniam}, \citenamefont {Narayanan},
		\citenamefont {Evers}, \citenamefont {Gruzberg},\ and\ \citenamefont
		{Mirlin}}]{mildenberger2007boundary}%
	\BibitemOpen
	\bibfield  {author} {\bibinfo {author} {\bibfnamefont {A.}~\bibnamefont
			{Mildenberger}}, \bibinfo {author} {\bibfnamefont {A.}~\bibnamefont
			{Subramaniam}}, \bibinfo {author} {\bibfnamefont {R.}~\bibnamefont
			{Narayanan}}, \bibinfo {author} {\bibfnamefont {F.}~\bibnamefont {Evers}},
		\bibinfo {author} {\bibfnamefont {I.}~\bibnamefont {Gruzberg}}, \ and\
		\bibinfo {author} {\bibfnamefont {A.}~\bibnamefont {Mirlin}},\ }\href@noop {}
	{\bibfield  {journal} {\bibinfo  {journal} {Phys. Rev. B}\ }\textbf {\bibinfo
			{volume} {75}},\ \bibinfo {pages} {094204} (\bibinfo {year}
		{2007})}\BibitemShut {NoStop}%
	\bibitem [{\citenamefont {Dybiec}\ \emph {et~al.}(2017)\citenamefont {Dybiec},
		\citenamefont {Gudowska-Nowak}, \citenamefont {Barkai},\ and\ \citenamefont
		{Dubkov}}]{dybiec2017levy}%
	\BibitemOpen
	\bibfield  {author} {\bibinfo {author} {\bibfnamefont {B.}~\bibnamefont
			{Dybiec}}, \bibinfo {author} {\bibfnamefont {E.}~\bibnamefont
			{Gudowska-Nowak}}, \bibinfo {author} {\bibfnamefont {E.}~\bibnamefont
			{Barkai}}, \ and\ \bibinfo {author} {\bibfnamefont {A.~A.}\ \bibnamefont
			{Dubkov}},\ }\href@noop {} {\bibfield  {journal} {\bibinfo  {journal}
			{Physical Review E}\ }\textbf {\bibinfo {volume} {95}},\ \bibinfo {pages}
		{052102} (\bibinfo {year} {2017})}\BibitemShut {NoStop}%
	\bibitem [{\citenamefont {Compte}(1996)}]{compte1996stochastic}%
	\BibitemOpen
	\bibfield  {author} {\bibinfo {author} {\bibfnamefont {A.}~\bibnamefont
			{Compte}},\ }\href@noop {} {\bibfield  {journal} {\bibinfo  {journal} {Phys.
				Rev. E}\ }\textbf {\bibinfo {volume} {53}},\ \bibinfo {pages} {4191}
		(\bibinfo {year} {1996})}\BibitemShut {NoStop}%
	\bibitem [{\citenamefont {Bay{\i}n}(2016)}]{bayin2016definition}%
	\BibitemOpen
	\bibfield  {author} {\bibinfo {author} {\bibfnamefont {S.~{\c{S}}.}\
			\bibnamefont {Bay{\i}n}},\ }\href@noop {} {\bibfield  {journal} {\bibinfo
			{journal} {Journal of Mathematical Physics}\ }\textbf {\bibinfo {volume}
			{57}} (\bibinfo {year} {2016})}\BibitemShut {NoStop}%
	\bibitem [{\citenamefont {Cai}\ and\ \citenamefont {Li}(2019)}]{cai2019riesz}%
	\BibitemOpen
	\bibfield  {author} {\bibinfo {author} {\bibfnamefont {M.}~\bibnamefont
			{Cai}}\ and\ \bibinfo {author} {\bibfnamefont {C.}~\bibnamefont {Li}},\
	}\href@noop {} {\bibfield  {journal} {\bibinfo  {journal} {Fractional
				Calculus and applied analysis}\ }\textbf {\bibinfo {volume} {22}},\ \bibinfo
		{pages} {287} (\bibinfo {year} {2019})}\BibitemShut {NoStop}%
	\bibitem [{\citenamefont {Lischke}\ \emph {et~al.}(2020)\citenamefont
		{Lischke}, \citenamefont {Pang}, \citenamefont {Gulian}, \citenamefont
		{Song}, \citenamefont {Glusa}, \citenamefont {Zheng}, \citenamefont {Mao},
		\citenamefont {Cai}, \citenamefont {Meerschaert}, \citenamefont {Ainsworth},\
		and\ \citenamefont {Karniadakis}}]{lischke2020fractional}%
	\BibitemOpen
	\bibfield  {author} {\bibinfo {author} {\bibfnamefont {A.}~\bibnamefont
			{Lischke}}, \bibinfo {author} {\bibfnamefont {G.}~\bibnamefont {Pang}},
		\bibinfo {author} {\bibfnamefont {M.}~\bibnamefont {Gulian}}, \bibinfo
		{author} {\bibfnamefont {F.}~\bibnamefont {Song}}, \bibinfo {author}
		{\bibfnamefont {C.}~\bibnamefont {Glusa}}, \bibinfo {author} {\bibfnamefont
			{X.}~\bibnamefont {Zheng}}, \bibinfo {author} {\bibfnamefont
			{Z.}~\bibnamefont {Mao}}, \bibinfo {author} {\bibfnamefont {W.}~\bibnamefont
			{Cai}}, \bibinfo {author} {\bibfnamefont {M.~M.}\ \bibnamefont
			{Meerschaert}}, \bibinfo {author} {\bibfnamefont {M.}~\bibnamefont
			{Ainsworth}}, \ and\ \bibinfo {author} {\bibfnamefont {G.~E.}\ \bibnamefont
			{Karniadakis}},\ }\href@noop {} {\bibfield  {journal} {\bibinfo  {journal}
			{Journal of Computational Physics}\ }\textbf {\bibinfo {volume} {404}},\
		\bibinfo {pages} {109009} (\bibinfo {year} {2020})}\BibitemShut {NoStop}%
	\bibitem [{\citenamefont {Baeumer}\ \emph {et~al.}(2018)\citenamefont
		{Baeumer}, \citenamefont {Kovács},\ and\ \citenamefont
		{Sankaranarayanan}}]{Baeumer2018}%
	\BibitemOpen
	\bibfield  {author} {\bibinfo {author} {\bibfnamefont {B.}~\bibnamefont
			{Baeumer}}, \bibinfo {author} {\bibfnamefont {M.}~\bibnamefont {Kovács}}, \
		and\ \bibinfo {author} {\bibfnamefont {H.}~\bibnamefont {Sankaranarayanan}},\
	}\href {\doibase https://doi.org/10.1016/j.jde.2017.09.040} {\bibfield
		{journal} {\bibinfo  {journal} {Journal of Differential Equations}\ }\textbf
		{\bibinfo {volume} {264}},\ \bibinfo {pages} {1377} (\bibinfo {year}
		{2018})}\BibitemShut {NoStop}%
	\bibitem [{\citenamefont {Kelly}\ \emph {et~al.}(2019)\citenamefont {Kelly},
		\citenamefont {Sankaranarayanan},\ and\ \citenamefont
		{Meerschaert}}]{Kelly2019}%
	\BibitemOpen
	\bibfield  {author} {\bibinfo {author} {\bibfnamefont {J.~F.}\ \bibnamefont
			{Kelly}}, \bibinfo {author} {\bibfnamefont {H.}~\bibnamefont
			{Sankaranarayanan}}, \ and\ \bibinfo {author} {\bibfnamefont {M.~M.}\
			\bibnamefont {Meerschaert}},\ }\href {\doibase
		https://doi.org/10.1016/j.jcp.2018.10.010} {\bibfield  {journal} {\bibinfo
			{journal} {Journal of Computational Physics}\ }\textbf {\bibinfo {volume}
			{376}},\ \bibinfo {pages} {1089} (\bibinfo {year} {2019})}\BibitemShut
	{NoStop}%
	\bibitem [{\citenamefont {Zoia}\ \emph {et~al.}(2007)\citenamefont {Zoia},
		\citenamefont {Rosso},\ and\ \citenamefont {Kardar}}]{Zoia2007}%
	\BibitemOpen
	\bibfield  {author} {\bibinfo {author} {\bibfnamefont {A.}~\bibnamefont
			{Zoia}}, \bibinfo {author} {\bibfnamefont {A.}~\bibnamefont {Rosso}}, \ and\
		\bibinfo {author} {\bibfnamefont {M.}~\bibnamefont {Kardar}},\ }\href
	{\doibase 10.1103/PhysRevE.76.021116} {\bibfield  {journal} {\bibinfo
			{journal} {Phys. Rev. E}\ }\textbf {\bibinfo {volume} {76}},\ \bibinfo
		{pages} {021116} (\bibinfo {year} {2007})}\BibitemShut {NoStop}%
	\bibitem [{\citenamefont {Garbaczewski}\ and\ \citenamefont
		{Stephanovich}(2019)}]{Garbaczewski2019}%
	\BibitemOpen
	\bibfield  {author} {\bibinfo {author} {\bibfnamefont {P.}~\bibnamefont
			{Garbaczewski}}\ and\ \bibinfo {author} {\bibfnamefont {V.}~\bibnamefont
			{Stephanovich}},\ }\href {\doibase 10.1103/PhysRevE.99.042126} {\bibfield
		{journal} {\bibinfo  {journal} {Phys. Rev. E}\ }\textbf {\bibinfo {volume}
			{99}},\ \bibinfo {pages} {042126} (\bibinfo {year} {2019})}\BibitemShut
	{NoStop}%
	\bibitem [{\citenamefont {Sokolov}\ \emph {et~al.}(2001)\citenamefont
		{Sokolov}, \citenamefont {Klafter},\ and\ \citenamefont
		{Blumen}}]{Sokolov2001}%
	\BibitemOpen
	\bibfield  {author} {\bibinfo {author} {\bibfnamefont {I.~M.}\ \bibnamefont
			{Sokolov}}, \bibinfo {author} {\bibfnamefont {J.}~\bibnamefont {Klafter}}, \
		and\ \bibinfo {author} {\bibfnamefont {A.}~\bibnamefont {Blumen}},\ }\href
	{\doibase 10.1103/PhysRevE.64.021107} {\bibfield  {journal} {\bibinfo
			{journal} {Phys. Rev. E}\ }\textbf {\bibinfo {volume} {64}},\ \bibinfo
		{pages} {021107} (\bibinfo {year} {2001})}\BibitemShut {NoStop}%
	\bibitem [{\citenamefont {Brockmann}\ and\ \citenamefont
		{Sokolov}(2002)}]{Brockmann2002}%
	\BibitemOpen
	\bibfield  {author} {\bibinfo {author} {\bibfnamefont {D.}~\bibnamefont
			{Brockmann}}\ and\ \bibinfo {author} {\bibfnamefont {I.}~\bibnamefont
			{Sokolov}},\ }\href {\doibase 10.1016/S0301-0104(02)00671-7} {\bibfield
		{journal} {\bibinfo  {journal} {Chemical Physics}\ }\textbf {\bibinfo
			{volume} {284}},\ \bibinfo {pages} {409} (\bibinfo {year}
		{2002})}\BibitemShut {NoStop}%
	\bibitem [{\citenamefont {Monroy}\ and\ \citenamefont
		{Raposo}(2024)}]{Monroy2024}%
	\BibitemOpen
	\bibfield  {author} {\bibinfo {author} {\bibfnamefont {D.~A.}\ \bibnamefont
			{Monroy}}\ and\ \bibinfo {author} {\bibfnamefont {E.~P.}\ \bibnamefont
			{Raposo}},\ }\href {\doibase 10.1103/PhysRevE.110.054119} {\bibfield
		{journal} {\bibinfo  {journal} {Phys. Rev. E}\ }\textbf {\bibinfo {volume}
			{110}},\ \bibinfo {pages} {054119} (\bibinfo {year} {2024})}\BibitemShut
	{NoStop}%
	\bibitem [{\citenamefont {Montroll}\ and\ \citenamefont
		{Scher}(1973)}]{Montroll1973}%
	\BibitemOpen
	\bibfield  {author} {\bibinfo {author} {\bibfnamefont {E.~W.}\ \bibnamefont
			{Montroll}}\ and\ \bibinfo {author} {\bibfnamefont {H.}~\bibnamefont
			{Scher}},\ }\href {\doibase https://doi.org/10.1007/BF01016843} {\bibfield
		{journal} {\bibinfo  {journal} {Journal of Statistical Physics}\ }\textbf
		{\bibinfo {volume} {9}},\ \bibinfo {pages} {101} (\bibinfo {year}
		{1973})}\BibitemShut {NoStop}%
	\bibitem [{\citenamefont {Angstmann}\ \emph
		{et~al.}(2015{\natexlab{a}})\citenamefont {Angstmann}, \citenamefont
		{Donnelly}, \citenamefont {Henry}, \citenamefont {Langlands},\ and\
		\citenamefont {Straka}}]{angstmann2015generalized}%
	\BibitemOpen
	\bibfield  {author} {\bibinfo {author} {\bibfnamefont {C.~N.}\ \bibnamefont
			{Angstmann}}, \bibinfo {author} {\bibfnamefont {I.~C.}\ \bibnamefont
			{Donnelly}}, \bibinfo {author} {\bibfnamefont {B.~I.}\ \bibnamefont {Henry}},
		\bibinfo {author} {\bibfnamefont {T.~A.~M.}\ \bibnamefont {Langlands}}, \
		and\ \bibinfo {author} {\bibfnamefont {P.}~\bibnamefont {Straka}},\
	}\href@noop {} {\bibfield  {journal} {\bibinfo  {journal} {SIAM Journal on
				Applied Mathematics}\ }\textbf {\bibinfo {volume} {75}},\ \bibinfo {pages}
		{1445} (\bibinfo {year} {2015}{\natexlab{a}})}\BibitemShut {NoStop}%
	\bibitem [{\citenamefont {Henry}\ \emph {et~al.}(2010)\citenamefont {Henry},
		\citenamefont {Langlands},\ and\ \citenamefont
		{Straka}}]{henry2010fractional}%
	\BibitemOpen
	\bibfield  {author} {\bibinfo {author} {\bibfnamefont {B.~I.}\ \bibnamefont
			{Henry}}, \bibinfo {author} {\bibfnamefont {T.}~\bibnamefont {Langlands}}, \
		and\ \bibinfo {author} {\bibfnamefont {P.}~\bibnamefont {Straka}},\
	}\href@noop {} {\bibfield  {journal} {\bibinfo  {journal} {Phys. Rev. Lett.}\
		}\textbf {\bibinfo {volume} {105}},\ \bibinfo {pages} {170602} (\bibinfo
		{year} {2010})}\BibitemShut {NoStop}%
	\bibitem [{\citenamefont {Sibuya}(1979)}]{sibuya1979generalized}%
	\BibitemOpen
	\bibfield  {author} {\bibinfo {author} {\bibfnamefont {M.}~\bibnamefont
			{Sibuya}},\ }\href@noop {} {\bibfield  {journal} {\bibinfo  {journal} {Annals
				of the Institute of Statistical Mathematics}\ }\textbf {\bibinfo {volume}
			{31}},\ \bibinfo {pages} {373} (\bibinfo {year} {1979})}\BibitemShut
	{NoStop}%
	\bibitem [{\citenamefont {Sibuya}\ and\ \citenamefont
		{Shimizu}(1981)}]{sibuya1981generalized}%
	\BibitemOpen
	\bibfield  {author} {\bibinfo {author} {\bibfnamefont {M.}~\bibnamefont
			{Sibuya}}\ and\ \bibinfo {author} {\bibfnamefont {R.}~\bibnamefont
			{Shimizu}},\ }\href@noop {} {\bibfield  {journal} {\bibinfo  {journal}
			{Annals of the Institute of Statistical Mathematics}\ }\textbf {\bibinfo
			{volume} {33}},\ \bibinfo {pages} {177} (\bibinfo {year} {1981})}\BibitemShut
	{NoStop}%
	\bibitem [{\citenamefont {Angstmann}\ \emph
		{et~al.}(2015{\natexlab{b}})\citenamefont {Angstmann}, \citenamefont
		{Donnelly}, \citenamefont {Henry},\ and\ \citenamefont
		{Nichols}}]{angstmann2015discrete}%
	\BibitemOpen
	\bibfield  {author} {\bibinfo {author} {\bibfnamefont {C.~N.}\ \bibnamefont
			{Angstmann}}, \bibinfo {author} {\bibfnamefont {I.~C.}\ \bibnamefont
			{Donnelly}}, \bibinfo {author} {\bibfnamefont {B.~I.}\ \bibnamefont {Henry}},
		\ and\ \bibinfo {author} {\bibfnamefont {J.~A.}\ \bibnamefont {Nichols}},\
	}\href@noop {} {\bibfield  {journal} {\bibinfo  {journal} {Journal of
				Computational Physics}\ }\textbf {\bibinfo {volume} {293}},\ \bibinfo {pages}
		{53} (\bibinfo {year} {2015}{\natexlab{b}})}\BibitemShut {NoStop}%
	\bibitem [{\citenamefont {Fogedby}(1994)}]{fogedby1994levy}%
	\BibitemOpen
	\bibfield  {author} {\bibinfo {author} {\bibfnamefont {H.~C.}\ \bibnamefont
			{Fogedby}},\ }\href@noop {} {\bibfield  {journal} {\bibinfo  {journal} {Phys.
				Rev. Lett.}\ }\textbf {\bibinfo {volume} {73}},\ \bibinfo {pages} {2517}
		(\bibinfo {year} {1994})}\BibitemShut {NoStop}%
	\bibitem [{\citenamefont {Fogedby}(1998)}]{fogedby1998levy}%
	\BibitemOpen
	\bibfield  {author} {\bibinfo {author} {\bibfnamefont {H.~C.}\ \bibnamefont
			{Fogedby}},\ }\href@noop {} {\bibfield  {journal} {\bibinfo  {journal} {Phys.
				Rev. E}\ }\textbf {\bibinfo {volume} {58}},\ \bibinfo {pages} {1690}
		(\bibinfo {year} {1998})}\BibitemShut {NoStop}%
	\bibitem [{\citenamefont {Jespersen}\ \emph {et~al.}(1999)\citenamefont
		{Jespersen}, \citenamefont {Metzler},\ and\ \citenamefont
		{Fogedby}}]{jespersen1999levy}%
	\BibitemOpen
	\bibfield  {author} {\bibinfo {author} {\bibfnamefont {S.}~\bibnamefont
			{Jespersen}}, \bibinfo {author} {\bibfnamefont {R.}~\bibnamefont {Metzler}},
		\ and\ \bibinfo {author} {\bibfnamefont {H.~C.}\ \bibnamefont {Fogedby}},\
	}\href@noop {} {\bibfield  {journal} {\bibinfo  {journal} {Phys. Rev. E}\
		}\textbf {\bibinfo {volume} {59}},\ \bibinfo {pages} {2736} (\bibinfo {year}
		{1999})}\BibitemShut {NoStop}%
	\bibitem [{\citenamefont {Meerschaert}\ and\ \citenamefont
		{Tadjeran}(2004)}]{meerschaert2004finite}%
	\BibitemOpen
	\bibfield  {author} {\bibinfo {author} {\bibfnamefont {M.~M.}\ \bibnamefont
			{Meerschaert}}\ and\ \bibinfo {author} {\bibfnamefont {C.}~\bibnamefont
			{Tadjeran}},\ }\href@noop {} {\bibfield  {journal} {\bibinfo  {journal}
			{Journal of computational and applied mathematics}\ }\textbf {\bibinfo
			{volume} {172}},\ \bibinfo {pages} {65} (\bibinfo {year} {2004})}\BibitemShut
	{NoStop}%
	\bibitem [{\citenamefont {Shen}\ \emph {et~al.}(2008)\citenamefont {Shen},
		\citenamefont {Liu}, \citenamefont {Anh},\ and\ \citenamefont
		{Turner}}]{shen2008fundamental}%
	\BibitemOpen
	\bibfield  {author} {\bibinfo {author} {\bibfnamefont {S.}~\bibnamefont
			{Shen}}, \bibinfo {author} {\bibfnamefont {F.}~\bibnamefont {Liu}}, \bibinfo
		{author} {\bibfnamefont {V.}~\bibnamefont {Anh}}, \ and\ \bibinfo {author}
		{\bibfnamefont {I.}~\bibnamefont {Turner}},\ }\href@noop {} {\bibfield
		{journal} {\bibinfo  {journal} {IMA Journal of Applied Mathematics}\ }\textbf
		{\bibinfo {volume} {73}},\ \bibinfo {pages} {850} (\bibinfo {year}
		{2008})}\BibitemShut {NoStop}%
	\bibitem [{\citenamefont {Wang}\ and\ \citenamefont
		{Barkai}(2020)}]{wang2020fractional}%
	\BibitemOpen
	\bibfield  {author} {\bibinfo {author} {\bibfnamefont {W.}~\bibnamefont
			{Wang}}\ and\ \bibinfo {author} {\bibfnamefont {E.}~\bibnamefont {Barkai}},\
	}\href@noop {} {\bibfield  {journal} {\bibinfo  {journal} {Physical Review
				Letters}\ }\textbf {\bibinfo {volume} {125}},\ \bibinfo {pages} {240606}
		(\bibinfo {year} {2020})}\BibitemShut {NoStop}%
	\bibitem [{\citenamefont {Hofert}(2011)}]{Hofert2011}%
	\BibitemOpen
	\bibfield  {author} {\bibinfo {author} {\bibfnamefont {M.}~\bibnamefont
			{Hofert}},\ }\href {\doibase https://doi.org/10.1016/j.csda.2010.04.025}
	{\bibfield  {journal} {\bibinfo  {journal} {Computational Statistics \& Data
				Analysis}\ }\textbf {\bibinfo {volume} {55}},\ \bibinfo {pages} {57}
		(\bibinfo {year} {2011})}\BibitemShut {NoStop}%
\end{thebibliography}
\end{document}